\documentstyle[12pt]{article}

\begin{document}
\begin{flushright}
IP/BBSR/98-29 \\
hep-ph/9809332 \\
\end{flushright}
\renewcommand{\thefootnote}{\fnsymbol{footnote}}
\begin{center}
{\LARGE STRANGENESS AS A QGP SIGNAL IN AN ISENTROPIC QUARK-HADRON  
PHASE TRANSITION}\vspace{.2in}\\
{\large B. K. Patra$^{1,}$\footnote[2]{E-mail: bkpatra@iopb.stpbh.soft.net} 
, K. K. Singh$^2$ and C. P. Singh$^2$ }\vspace{.2in}\\
{\it {$^1$}Institute of Physics, Sachivalaya Marg, Bhubaneswar 751 005, India\\
$^2$Department of Physics, Banaras Hindu University, 
Varanasi 221 005, India.}\vspace{.2in}\\
\underline{Abstract}
\end{center}
Lattice QCD results reveal that the critical parameters and the order of the 
quark-hadron phase transition are quite sensitive to the number of dynamical
flavours and their masses included in the theory. Motivated by this
result we develop a phenomenological equation of state for the quark-gluon
plasma consisting of $n_f$ flavours retaining the entropy per baryon ratio
continuous across the quark-hadron phase boundary. We thus obtain a generalised
expression for the temperature and baryon chemical potential dependent bag
constant. The results are shown for the realistic case, i.e., involving u, d
and s quarks only. We then obtain a phase boundary for an isentropic 
quark-hadron
phase transition using Gibbs' criteria. Similarly another phase boundary
is obtained for the transition to an ideal QGP from the solution of the 
condition  $B(\mu,T)=0$. The
variation of critical temperature $T_c$ with the number of flavours included
in the theory. Also the variation of ${(\varepsilon-4P)}/T^4$ with temperature 
are studied
and compared with lattice results. Finally the strange particle ratios
$\frac{\overline{\Lambda}}{\Lambda}$, $\frac{\overline{\Xi}}{\Xi}$ and 
$\frac{K^+}{K^-}$ are obtained 
at both
phase boundaries. We propose that their variations with the temperature
and baryon chemical potential can be used in identifying the quark-gluon
plasma in the recent as well as in future heavy-ion experiments.\\

\pagebreak
\section{\bf Introduction}
Recent lattice gauge simulations reveal interesting results [1-3] for the
number of flavours and their masses included in the theory, in determining the
order and the critical parameters of the phase transition. In realistic
cases, i.e., for three or more flavours, the transition is found to be a first 
order. However, for two light flavours, one finds a continuous transition.
As the mass of the strange quark is increased from zero, the transition changes
from first-order to a continuous one. Similarly the critical parameters are
also sensitive to the number of flavours $n_f$ and it is shown
that the critical temperature $T_c$ drops with increasing flavours
($T_c$ $\propto$ ${n_f}^{-1/2}$). The lattice approach, however, is still
not a suitable tool for studying the strongly interacting matter in the
baryon rich environment. Therefore, the critical behaviour of such a
matter with finite baryon chemical potential can reliably be studied
in the framework of a QCD-motivated  phenomenological models. However,
within the framework of such models, the specific entropy per baryon
($S/B$) is found to be discontinuous across the phase boundary [4-6].
Leonidov et al. [7] have recently proposed a modified equation of state (EOS)
for the quark-gluon plasma (QGP) with a temperature T and baryon chemical
potential $\mu$ dependent bag constant in order to ensure that the $S/B$
ratio is continuous along the phase boundary. In our earlier works [8-9],
 we have modified the $\mu$ and T dependence of B($\mu$,T) by incorporating
the QCD perturbative corrections in the EOS for QGP and we have also
explored in detail the consequences of such a bag constant on the
deconfining phase transition. The purpose of this paper is to generalise
the expression of B($\mu$,T) for $n_f$ flavours in the low baryon
density region and then we investigate the effect of a massless as well as
a massive s-quark on the critical behaviour of the phase transition.
We compare our results with those of lattice in the high T region
qualitatively. It should be noted that when $S/B$ ratio is made continuous
across the phase boundary, the quark-hadron phase transition is still
first-order and it occurs at a common temperature and chemical potential.\\  
We have then determined the critical parameters from the two phase boundaries
 - one obtained from the Gibbs criteria for an isentropic, equilibrium
phase transition and the other for the transition to an ideal, noninteracting
QGP. Finally we study the variations of the ratios
$\frac{\overline{\Lambda}}{\Lambda}$, $\frac{\overline{\Xi}}{\Xi}$ and 
$\frac{K^+}{K^-}$  
either with the temperature or with baryon chemical potential at
both phase boundaries. Here we  mainly concentrate on the particle 
production from the QGP in the midrapidity region so that it can be
used for studying the properties of QGP produced at future RHIC and LHC
experiments.\\
\vspace{0.120 in}
\section{\bf  Formulation of B($\mu$,T) for $n_f$ Flavours}
The chemical potential ($\mu$) and temperature dependence ($T$) of the bag
constant can be derived by constructing an isentropic, equilibrium phase
transition from a quark-gluon plasma (QGP) to a hadron gas (HG) at a fixed
T and $\mu$, and we put a constraint that entropy per baryon ratio is
continuous across the phase boundary, i.e.,
\begin{eqnarray} 
{\left. {\frac{S}{B}}\right|}_{QGP} = {\left. {\frac{S}{B}}\right|}_{HG}
\end{eqnarray} 
The above condition can be achieved by assigning a T and $\mu$ dependence
to the bag constant explicitly [7]. The resulting correction factors
$- \frac{\partial B(\mu,T)}{\partial T}$ and 
$- \frac{\partial B(\mu,T)}{\partial \mu}$ to the entropy density and the
baryon density, respectively, in the QGP phase will modify eq.(1) as
follows : 
\begin{eqnarray} 
\frac{S^{o}_{QGP} - \frac{\partial B(\mu, T)}{\partial T}}
{n^{o}_{QGP} - \frac{\partial B(\mu, T)}{\partial \mu }}
 = \frac{S_{HG}}{n_{HG}}
\end{eqnarray} 
The above partial differential equation can be solved iteratively by
exploiting the symmetry property of the bag pressure, i.e., $B(\mu,T) = 
B(-\mu,-T)$.\\
We use simple models to describe the EOS in each of the two phases and then
perform a Maxwell construction to determine their common phase boundary. The
QGP phase consists of a perturbatively interacting gas of quarks and gluons.
Hence the thermodynamical potential for a QGP can be written in terms of the
grand partition function as[10]
\begin{eqnarray}
\Omega(T,V,\mu) = - \frac{T}{V} \ln Z_{QGP}
\end{eqnarray}
and $\ln Z_{QGP}$ = $\ln {Z^o}_{QGP}$ + $\ln {Z^{Int}}_{QGP}$ + $\ln {Z^{Vac}}_
{QGP}$, where $Z^o$ is the zeroth order contribution, $Z^{Int}$ arises due to
perturbative corrections and the term $Z^{Vac}$ represents the
non-perturbative vacuum contribution in the form of a $\mu$ and T dependent
bag constant - B($\mu$,T).\\
The zeroth order contribution to the partition function is :
\begin{eqnarray}
\ln Z^o_{QGP} =  - \frac{V}{T} 
\sum_f \frac{g_f}{6 {\pi}^2} \int_{0}^{\infty}\! \frac{p^4  dp}{\sqrt{(p^2 + {m_f}^2)}} 
~ F_f[p; T, {\mu}_f] 
\end{eqnarray}
and the perturbative interaction part up to second order in strong coupling
constant $g_s$ (${\alpha}_s$ = $\frac{{g_s}^2}{4 \pi})$ can be written as
follows by using finite temperature field theory :
\begin{eqnarray}
\ln Z^{Int}_{QGP} =  - \frac{V}{T} \left[ \frac{1}{3}~ \pi~ {\alpha}_s~ N_g 
~T^2 \sum_f \int_{0}^{\infty}\! \frac{d^3p}{{(2\pi)}^3}  \frac{1}{E_p}
 F_f[p; T, {\mu}_f]  \right. ~\rm~~  \nonumber\\
+ \pi {\alpha}_s N_g \sum_f \int_{0}^{\infty}\!\int_{0}^{\infty}\! 
\frac{d^3p}{{(2\pi)}^3}
\frac{d^3q}{{(2\pi)}^3} \frac{[{F_f^-}(q){F_f^-}(p) + 
{F_f^+}(q){F_f^+}(p)]}{{E_p}{E_q}} 
\left( \frac{2 {m_f^2}} 
{{(E_p - E_q)}^2 - {\omega}^2} + 1 \right)  \nonumber \\
+ \pi {\alpha}_s N_g \sum_f \int_{0}^{\infty}\!\int_{0}^{\infty}\! \frac{d^3p}{{(2\pi)}^3}
\frac{d^3q}{{(2\pi)}^3} \frac{[{F_f^-}(q){F_f^+}(p) + 
{F_f^+}(q){F_f^+}(p)]}{{E_p}{E_q}} \left( \frac{2 {m_f^2}} 
{{(E_p + E_q)}^2 - {\omega}^2} + 1 \right)  \nonumber \\ 
+ \left. \frac{1}{36}~\pi~{\alpha}_s~{N_c}~{N_g}~{T^4}  
 \right]    
\end{eqnarray} 
where $F_f$ =$ F_f^+$ +$ F_f^-$; $ F_f^+  (F_f^-)$ are the fermionic 
(antifermionic) equilibrium distribution functions of f-th flavour, 
respectively :

\begin{eqnarray}
F_f^+ = {\left[ {e^{~\beta(E_f - \mu_f)} + 1}
\right] }^{-1} 
\end{eqnarray}
\begin{eqnarray}
F_f^- = {\left[ {e^{~\beta(E_f + \mu_f)} + 1} \right]}^{-1} 
\end{eqnarray}

Here $E_f$ is the energy of f-flavour. Considering massless quarks the total
thermodynamic potential for the QGP phase reduces to the following expression
[11] :
\begin{eqnarray}
\Omega = - \frac{\pi^2}{45}~ T^4 \left[ 8 + \frac{21}{4} ~ n_f + \frac{45}
{2 \pi^2} \sum_f g({\mu}_f,T) \right] \nonumber \\
+ \frac{2 \pi}{9}~{\alpha}_s ~T^4 \left[ 3 + \frac{5}{4} ~ n_f + \frac{9}
{2 \pi^2} \sum_f g({\mu}_f,T) \right] + B({\mu}_f,T)
\end{eqnarray} 
where $g(\mu_f,T)$ = $ \left[ \frac{\mu_f^2}{T^2} + \frac{\mu_f^4}{2 \pi^2 T^4} \right]$
\vspace{.1in} \\
If we further make use of the various conservation conditions such as
strangeness ($\mu_s$ = 0) and charm conservations, and also put
$ \mu_u $ = $ \mu_d $ = $ \mu_q $ = $ \frac{\mu}{3} $, we get the
expression for the pressure in QGP :
\begin{eqnarray}
P_{QGP} = \frac{\pi^2}{45}~ T^4 \left[ 8 + \frac{21}{4} ~ n_f + \frac{45}
{2 \pi^2} \sum_q g({\mu}_q,T) \right] \nonumber \\
- \frac{2 \pi}{9}~{\alpha}_s ~T^4 \left[ 3 + \frac{5}{4} ~ n_f + \frac{9}
{2 \pi^2} \sum_q g({\mu}_q,T) \right] - B({\mu}_q,T)
\end{eqnarray} 
In these calculations, $\mu$ and T dependence of $\alpha_s$ is taken as
follows [12] :
\begin{eqnarray}
\alpha_s (\mu_f , T) = \frac{4 \pi} {(11 - \frac{2}{3} n_f) \ln \frac{M^2}
{\Lambda^2}}
\end{eqnarray}
where
\[
M^2 = \frac{4}{3} \left [ 16 \int_{0}^{\infty}\! dp~p^4 f_G + \sum_f g_f 
\int_{0}^{\infty}\! dp~p^4 F_f 
\right] {\left [ 16 \int_{0}^{\infty}\! dp~p^4 f_G + \sum_f g_f \int_{0}^
{\infty}\! dp~p^4 F_f \right]}^{-1} 
\]
Here $f_G$ is the gluonic and $F_f$ is the quark distribution functions,
$g_f$ is the quark degeneracy factor for the f-th flavour.\vspace{.1in}\\
Hadron gas consists of all strange as well as nonstrange hadrons. The
partition function for baryons (B) then becomes
\begin{eqnarray}
\frac{T}{V} \ln Z_B (T,\mu) & = & \left[  \sum_{i=1}^{n_1} \frac{g_i m_i^2 T^2}
{\pi^2} ~ K_2 (\frac{m_i}{T}) \right] \cosh(\frac{\mu}{T}) \nonumber \\
& + & \left[  \sum_{j=1}^{n_2} \frac{g_j m_j^2 T^2}
{\pi^2}~K_2 (\frac{m_j}{T}) \right] \cosh(c \frac{\mu}{T}) \nonumber \\
& + & \left[ \sum_{k=1}^{n_3} \frac{g_k m_k^2 T^2}
{\pi^2} K_2 (\frac{m_k}{T}) \right] \cosh(d \frac{\mu}{T}) 
\end{eqnarray}
here c = 1 - $f_s$ and d = 1 - 2$f_s$ and $f_s(\equiv \frac{\mu_s}{\mu})$
is a  factor arising from the strangeness conservation condition $n_s$ -
$ n_{\overline s}$ = 0 in the hadronic phase and $n_1$, $n_2$ and $n_3$
are the number of nonstrange, singly and doubly strange baryons, respectively.
The summation arises due to quantum statistics used for the distribution
functions. Similarly the partition function
for the mesonic (M) sector is
\begin{eqnarray}
\frac{T}{V} \ln Z_M (T,\mu) & = & \frac{\pi^2}{30} T^4  + \sum_{l=1}^{r_1} 
\frac{g_l m_l^2 T^2} {\pi^2}~K_2 (\frac{m_l}{T}) \nonumber \\
& + &  \sum_{n=1}^{r_2} \frac{g_n m_n^2 T^2}
{\pi^2}~K_2 (\frac{m_n}{T})  \cosh( \frac{f_s}{T}) 
\end{eqnarray}
Here the first term is for massless pions, second and third term is
due to nonstrange and strange mesons, respectively. In order to
incorporate repulsive interactions among the hadrons in the HG phase
within the excluded-volume approach, we multiply all pointlike
quantities by a volume correction factor [13] of the type
[ 1 + $V_o$ $n_{HG}^o$(T,$\mu$)]$^{-1}$ where $n_{HG}^o$(T,$\mu$) is the
net baryon density for pointlike baryons.\vspace{.15in}\\
The requirement of conserved entropy per baryon across the phase boundary
leads to the following expression in the low baryon density
and high temperature limit :
\begin{eqnarray}
\frac{\partial B}{\partial \mu} \simeq n_{QGP}^o - \frac{S_{QGP}^o}{S_M} 
~n_{HG}
\end{eqnarray}

We have neglected the terms proportional to $\mu^2$ and $\mu^3$ as well as
$\frac{\partial B}{\partial T}$ term. We have also assumed that 
the entropy
in the hadron gas at a large T and very small $\mu$ is mainly contributed by 
mesons and thus $S_{HG}$ $\simeq$ $S_M$.\\
In the low baryon density limit, we can approximate the entropy density 
$S_{QGP}^o$ as almost independent of $\mu$ and hence we get,

\begin{eqnarray}
B(\mu,T) = B_o + \frac{1}{9}~\mu^2 ~T^2 - \frac{S_{QGP}^o}{S_M} ~ 
\Big[
P_B(T,\mu) - P_B(T, \mu \simeq 0) \Big] 
\end{eqnarray}
Here $P_B$ is the baryonic pressure in the HG phase. Although $S_{QGP}^o$
may not be $\mu$-independent but it involves $\mu^2$ and higher order
terms, which we ignore in the $\mu$ $\rightarrow$ 0 limit. Finally after
second iteration, we get the expression for B($\mu$,T) at high T
and small $\mu$ limit as :
\begin{eqnarray}
B(\mu,T) & = & B_o + \frac{1}{9} ~ \mu^2 ~T^2 + \frac{\mu^4}{162 \pi^2} 
- \frac{S_{QGP}^o}{S_M} \Big[ P_B(T,\mu) - P_B(T, \mu \simeq 0) \Big] 
\nonumber \\
& - & \frac{1}{2} \left( \frac{S_{QGP}^{o^\prime}}{S_M} - S_{QGP}^o  \frac
{S_M^\prime}{S_M^2} \right) {\Big( P_B(T,\mu) - P_B(T, \mu \simeq 0)
\Big)}^2
\end{eqnarray}
Here
\[
S_{QGP}^{o^\prime} = \frac{\partial S_{QGP}^o}{\partial T}  ~{\rm and}~ 
  S_M^\prime = \frac{\partial S_M}{\partial T} 
\]
Using these equations of state (EOS) for the QGP and HG, one can
construct a first-order isentropic and equilibrium phase transition with the help
of Gibbs criteria i.e.,  $P_{QGP}$ = $P_{HG}$ and thus determine the
critical parameters $T_c$, $\mu_c$ for the phase boundary. Similarly
one can also determine the other phase boundary by determining
$T_{c^\prime}$, $\mu_{c^\prime}$ for the phase transition from a hadron gas
to an ideal QGP from the condition  B($\mu$,T) = 0 in the eq.(15).
We can then calculate the strange particle ratios along these
boundaries. We choose $T_{c^\prime}$ as the approximate temperature
where ${(\varepsilon - 3P)}/T^4$ = 0, signifying the onset of the
ideal QGP behaviour ($\varepsilon$ and P being the energy density
and pressure in the QGP phase, respectively). Thus we get two sets of critical
parameters ($T_c$, $\mu_c$) and ($T_{c^\prime}$, $\mu_{c^\prime}$) along
which we will calculate strange particle ratios. The region under the
curve defined by ($T_c$, $\mu_c$) is considered effectively as the hadron gas
phase, while the ideal QGP phase resides above the curve defined by
($T_{c^\prime}$, $\mu_{c^\prime}$). The region between the two curves
may be considered as the transition region from a deconfined phase
consisting of massive quarks into a chiral symmetry restored ideal
QGP phase. The strange particle ratios along ($T_c$, $\mu_c$) curve
predict quantitavely the minimum QGP values because they represent
the deconfining phase boundary. Similarly ($T_{c^\prime}$, $\mu_{c^\prime}$)
curve ensures that, if measurements of the observables yield
an experimental value for temperature and quark chemical potential
above this curve, an ideal quark-gluon plasma has been produced
in the heavy-ion collisions.\\
\section{\bf Calculation of Strange particle Ratios}
The idea that the strangeness abundance can provide a useful
signature for quark-gluon plasma formation originates from
the enhancement of the strange-antistrange quark ($s\overline s$)
pairs in a thermally equilibrated and baryon dense QGP as compared to the
production of u, d light quark flavours[14]. This is possible only
when the Fermi energy of the u, d quarks become larger than the strange
quark mass and hence the Pauli blocking stops further creation of
light quark pairs. For a $q\overline q$ symmetric and baryonless QGP,
the abundance of $s\overline s$ can still occur provided the temperature
is much larger than the strange quark mass. Furthermore, the strangeness
abundance in QGP is also related with the lower mass threshold involved
in the creation of an $s\overline s$ pair in QGP rather than $K\overline K$
pair in the hadron gas. Thus it seems worthwhile to investigate
in detail the utility of strangeness as a QGP signal in the context
of future LHC and RHIC experiments where the production of baryonless
QGP is expected. However, in this context, we will be searching for the
strange particle ratios $\frac{\overline \Lambda}{\Lambda}$,
$\frac{\overline\Xi}{\Xi}$ and $\frac{K^+}{K^-}$ as a QGP signal.
\vspace{.1in}\\
In calculating the above ratios we entirely dwell on the formalism
in the quark-gluon plasma phase formed at a high temperature
and a small baryon density. Thus the composition of the QGP mainly
involves the gluons and hence the glue-based processes dominantly
contribute to the strangeness production. The probability for the
creation of quarks of different flavours is calculated in the particle 
production mechanism [15,16] :
\begin{eqnarray}
f_q = f_o ~~ e^{ - \frac{\pi m_q^2}{\kappa}}
\end{eqnarray}
where $m_u$ = 5 MeV, $m_d$ = 9 MeV, $m_s$ = 170 MeV,  and the QCD
string tension $\kappa$ = 1 ($GeV/fm$). The normalisation constant $f_o$
is given by
\begin{eqnarray}
f_o = \frac{1}{ 2 + e^{ - \frac{\pi m_s^2}{\kappa}}}
\end{eqnarray}
The hadronisation of QGP is still not understood properly. We
assume that once QGP has been produced, the resulting hadrons do not
get sufficient time to achieve thermal and chemical equilibrium. Thus 
the chemical composition existing during the QGP stage does not change
and hence the strangeness abundance achieved in the QGP phase still
survives during the process of hadronisation. Thus the probability for the creation
of a primordial hadron with q number of quarks per unit of phase space volume
is obtained
\begin{eqnarray}
P = \prod_q ~  f_q ~  \lambda_q  ~ g_q  ~ \gamma_q ~ e^{ - \frac{E_q}{T}}
\end{eqnarray}
where $f_q$ is the gluon fragmentation probability, and involves the quark 
mass dependence, the quark fugacity $\lambda_q $ = $e^{ \mu /T}$,
$g_q$ is the statistical degeneracy factor, $\gamma_q$ is the relative
equilibration factor and $E_q$ is the energy of the q-th quark. We
assume complete equilibration for u, d, s quarks in QGP, i.e, 
$\gamma_u$ = $\gamma_d$ =  $\gamma_s$ = 1. The  energy of the emitted
particle is E = $\sum_q E_q$. This factor is integrated out in eq.(18)
when we carry out the phase space integration and take the particle ratios
in the same $m_T$ range : E = $m_T \cosh (y_{pr} - y_{mr})$, where $m_T$ = 
$\sqrt{m^2 + p_T^2}$ [18], with $y_{pr}$ ($y_{mr}$) being the projectile (mid)
rapidity. Thus for ratios like $\frac{\overline \Lambda}{\Lambda}$,
$\frac{\overline \Xi}{\Xi}$ and $\frac{K^+}{K^-}$, the numerator as well
as denominator gives similar kinematical factor after integrations and hence
they are cancelled out. However, these ratios depend upon the product
of $f_q ~\lambda_q $ and hence on the mass of the quark as well as  on the
temperature and quark chemical potential. We introduce  $\lambda_i $ = 
$e^{ \mu_i  /T}$ as the fugacity of i-th hadron species in HG  and is simply
the product of the fugacities of the constituent quarks so that $\lambda_N$ = 
$\lambda_q^3$, $\lambda_K$ = $ \lambda_q \lambda_{\overline s} $, etc. 
Since the u, d, s flavours are separately conserved in the time scale of hadronic
collisions, the production (or annihilation) can only
occur in pairs. So the chemical
potentials for particle and antiparticle are opposite to each other, we get
$\lambda_q$ = $\lambda_q^{-1}$. Finally, particle ratios in the same
$m_T$ range, take the form [17] :
\begin{eqnarray}
\frac{\overline \Lambda}{\Lambda} & = & e^{- 2(\mu_u + \mu_d)/T}
~~ e^{- 2 \mu_s/T} \\
\frac{\overline \Xi}{\Xi} & = & e^{- 2\mu_u/T}
e^{- 4\mu_s/T} \\
\frac{K^+}{K^-} & = & e^{-  2\mu_u/T}
e^{- 2\mu_s/T} 
\end{eqnarray}
All the above ratios depend on the factor ${\it exp}{(\pm \mu_q/T)}$ and
strange quark chemical potential $(\mu_s)$ explicitly. In the pure QGP 
phase, however,
$\mu_s$ is identical to zero because of the exact strangeness
conservation. The factor $\gamma_s$ accounts for much of our ignorance
about the dynamics of strangeness formation and the aproach to 
equilibration of the strange quarks in the QGP phase. A value $\gamma_s$ = 1 is 
believed to favour QGP interpretation of the data. It is convenient to denote  
$\mu_q$ = $(\mu_u + \mu_d)$/2; $\delta\mu$ = $ \mu_d - \mu_u$, 
$\mu_B$ = 3 $\mu_q$ ; where $\mu_q$ is quark chemical potential,
$\mu_B$ is the baryochemical potential and $\delta \mu$ describes the
(small) asymmetry in the number of up and down quarks due to neutron 
excess in heavy-ion collisions, i.e., $\frac{Z}{A} \neq $ 0.5. In practice
we find $\delta \mu$ to be very small [19], and, therefore, we neglect 
it here. Finally we can calculate these ratios on the
phase boundaries using the values of the critical parameters
($T_c$, $\mu_c$) and ($T_{c^\prime}$, $\mu_{c^\prime}$).\\
\section{\bf Results and Discussions}
In Fig.1 we have shown the variation of $B(\mu,T)$  with temperature
T as obtained from eq.(15) in the low $\mu$ and high T limit. We find
that the values for B($\mu$,T) differ much from the case of u, d massless
quarks and decreases faster as T increases. Moreover, the incorporation
of interactions into the calculation modifies the result significantly.
We find that B($\mu$,T) becomes zero for T = 210 MeV if the interactions
among quarks and gluons are incorporated and it is far less than the value T = 440 MeV
obtained in the case for three massless quarks included in the QGP without
any interactions as shown by curve B. Vanishing of B($\mu$,T) at a certain
value of $\mu$ and T signifies that the quarks and gluons are almost free.
We infer that the ideal gas limit for a QGP consisting of three
flavours is reached at a value $T \gg T_c $ with the critical temperature
$T_c \simeq $ 160 MeV for the deconfining phase transition. However, we find
that the calculation shows very little change when we take u, d massless
quarks and s as  a massive quark ($m_s$ = 150 MeV) if compared with the case
of three massless quarks. \vspace{.1in}\\
In order to illustrate the above point more clearly, we have plotted
in Fig.2 the variation of the quantity $(\varepsilon - 3P)/T^4$
with temperature (T) for a QGP consisting of u, d, s massless quarks. Here
$\varepsilon$ represents the energy density and P is the pressure of the
QGP. The quantity $(\varepsilon - 3P)$ represents a measure of the ideal gas
behaviour since it vanishes for an ideal gas. We find that
$(\varepsilon - 3P)/T^4$ asymptotically vanishes. This curve agrees
with that obtained in the lattice gauge results. Thus it
gives us confidence in the QCD-motivated calculations for the EOS of a QGP.
\vspace{.1in}\\
We conclude that the inclusion of $n_f$ flavours in the
temperature and baryon chemical potential dependent bag constant yields
results in agreement with those obtained in the lattice gauge results [3].
However, we find that there is very little difference in the results
whether the strange quark is massless or massive.
This result differs from the lattice finding. Moreover, our phenomenological
model is valid for a QGP with a finite but small baryon chemical potential.
The B($\mu$,T) thus obtained guarantees the phase transition between
 QGP and hadron gas at the same value of temperature and chemical
potential and hence the process of reheating is not required during
the mixed phase of hadronisation. It also yields continuity of entropy per baryon
ratio across the phase boundary. We thus hope that the phenomenological
form for B($\mu$,T) thus obtained will be of considerable
use in deriving the properties and the signals of QGP to be produced
at proposed RHIC and LHC experiments because we expect the net baryon density in these 
experiments to be low enough as used in the above calculation. However,
the baryon density will never vanish precisely and, therefore, the lattice
calculations can still not be used as such here. We, therefore, hope
that the description of a deconfinement phase transition achieved in a QCD-motivated
model by making the bag constant dependent on $\mu$ and T will be
a more realistic one in the future LHC and RHIC experimental situations.
\vspace{.1in}\\
In Fig.3, we have separately shown the variations of  $\frac{\overline 
\Lambda}{\Lambda}$  with the baryon chemical potential and also with 
the temperature.
We have separated the regions of HG phase, interacting QGP phase and the
ideal or non-interacting QGP phase. Our results are valid in the low
baryon density limit and hence cannot be extended beyond $\mu > $ 300 MeV
which corresponds to $\mu_q$ = 100 MeV. In future colliders like RHIC
and LHC, we expect the formation of an ideal quark matter at $\mu \leq
$ 100 MeV. It means that the value of our calculated ratio $\frac{\overline \Lambda}
{\Lambda} \geq $ 0.56 should indicate the formation of such a matter.
However, the variation of  $\frac{\overline \Lambda}{\Lambda}$ with the
temperature does not show any similar constraint on the temperature. 
It simply tells us
that an ideal QGP can be formed at temperatures T $<$ 180 MeV. The
present experimental value [20,21] of  $\frac{\overline \Lambda}{\Lambda}$
is 0.22 $\pm$ 0.01 and it signifies a hot and dense HG rather
than a QGP unless $\mu > $ 300 MeV for which our predictions are rather
unreliable and inconclusive. \vspace{.1in}\\
In Fig.4, we have shown the similar variations for the ratio $\frac{\overline
\Xi}{\Xi}$ with $\mu$ and T separately. We again find that an extraordinary large
value for this ratio ( $>$ 0.74) at $\mu < $ 100 MeV will indicate the
formation of QGP at RHIC or LHC experiments. In Fig.5, we have also shown
the variations of $\frac{K^+}{K^-}$ either with $\mu$ or with T. In
this case, we find that the formation of an ideal QGP at $\mu < $ 100 MeV
requires the value of the ratio  $\frac{K^+}{K^-}$ $\leq$ 1.3. \vspace{.1in}\\
Recently Asprouli and Panagiotou [17] performed an identical
analysis. However, they fixed the phase boundaries
by choosing $T_{c^\prime}$, $\mu_{c^\prime}$ etc. quite arbitrarliy. Thus their comparison
with the experimental data revealed a faulty conclusion that the present
CERN and AGS heavy-ion experimental data [20,21] indicate QGP formation.
Furthermore, they conclude that Sulphur induced reactions at midrapidity
region reveal that the ideal QGP has been produced to within 60 \% 
possibility.
Our results, on the contrary, suggest that the CERN experimental datas
correspond to an ideal thermalised HG picture. However, the results
obtained here clearly indicate that strange particle ratios can provide a
signal in the midrapidity and  baryon-free region. The experimental results
do not correspond to this region. Nevertheless, the EOS
employed here for the QGP phase breaks down strictly at $n_B$=0 region
because entropy per baryon ratio becomes meaningless. So our analysis
is applicable only in the low baryon density ($n_B$) region. \vspace{.1in}\\
It has been claimed that thermal gluon decay into a quark-antiquark pair
dominates for a wide range of quark masses in the midrapidity region
and thus provides the most important source for the quark or the
antiquark density [22,23]. Normally, the gluon cannot decay
into a strange quark-antiquark pair because its thermal mass is below the
threshold required for such a pair creation. For a temperature (T) around
200 MeV and the coupling constant g = 2 for a QGP consisting of two massless 
flavours, one gets the thermal gluon mass 
$m_g$ = $\frac{2}{3}$
g T = 267 MeV in the lowest order perturbation theory and threshold  for 
$s\overline s$ pair creation 
 corresponds to $\simeq$ 300 MeV. However, it has been suggested
that in addition to acquiring a thermal mass of the order of $g^2$T,  gluons
also acquire a width determined by the large damping rate [24,25]. Thus g $\rightarrow$ $s \overline s$ decay is allowed and thus can account
for the strangeness enhancement in the midrapidity region. \vspace{.15in}\\
In summary, we have examined the strange particle production in the 
midrapidity region within a QGP formalism. We assume that the gluon 
fragmentation into $ q\overline q$ pairs is the main source of the quark
and antiquark density in the plasma. However, we have calculated the
values of the critical parameters for the transition from the HG to an
ideal QGP using our EOS for the quark matter
with a $\mu$ and T dependent bag constant obtained at a low baryon
density. The lower limit to the critical parameters corresponds to Gibbs
criteria for an equilibrium and isentropic phase transition and upper
limit is obtained by putting $B(\mu,T)$ = 0. We have plotted the
strange particle ratios as a function of T and $\mu$, separately along
the two phase curves. From the values of these strange particle ratios 
one can infer 
an approximate values of temperature and the baryon chemical potential
reached in the heavy-ion experiments. Our studies reveal that the strangeness
enhancement can still be regarded as the signature for a QGP formed
in the central rapidity or at almost baryon-free region. However, in the absence of a reliable
lattice result for $n_B$ $\neq$ 0 region, our approach
is justified. We hope that our results will provide a basis for
future studies regarding the signals of QGP and their detection
in the future collider experiments.\\
\pagebreak

\newpage
\noindent {\large \underline{Figure Captions}}\vspace{.5in}\\
Fig. 1. Variation of bag constant $B(\mu,T)$ with temperature (T) at a
baryon chemical potential $\mu$ = 50 MeV. Curve A represents the free QGP
EOS consisting of u, d massless flavours. Curve B is for the free u, d, s
massless flavours. Curve C stands for u, d massless and interacting flavours
with QCD scale parameter $\Lambda $ = 100 MeV. Curve D represents interacting
QGP EOS but s-quark is massive ( $m_s$ = 150 MeV). Curve E is same as
D but s-quark is massless. In all these curves, we have used $B_o^{1/4}$ = 
235 MeV.  \vspace{.2in}\\

\noindent Fig. 2. Variation of $(\varepsilon - 3P)/T^4$ which is a measure
of an ideal plasma behaviour with temperature (T) for $\mu$ = 50 MeV and
$B_o^{1/4}$ = 235 MeV. Here $\varepsilon$ is the energy density and P is the
pressure of the QGP phase of three massless flavours.\vspace{.2in}\\ 
\noindent Fig. 3. Variations of $\frac{\overline \Lambda}{\Lambda}$ with
$\mu$ and T are shown separately. The dashed line represents the ratio
on the phase boundary determined by Gibbs criteria for equilibrium phase
transition between interacting QGP and HG whereas the solid line denotes
the ratio on the phase boundary between HG and ideal QGP phases.\vspace{.2in}\\
\noindent Fig. 4. The notations are the same as Fig. 3 but for the ratio 
$\frac{\overline \Xi}{\Xi}$. \vspace{.2in}\\ 
\noindent Fig. 5. The notations are the same as Fig. 3 but for the ratio 
$\frac{K^+}{K^-}$. \vspace{.2in}\\ 

\end{document}